\documentclass[aps,preprint,showpacs,amsmath,amssymb,floatfix]{revtex4-1}

\usepackage{graphicx}
\usepackage{dcolumn}
\usepackage{bm}
\usepackage[T1]{fontenc}

\usepackage{color}

\begin{document}

\title{Neutral-current neutrino-nucleus cross sections based on relativistic nuclear energy density functional}


\author{H. {\DJ}apo$^{1,2}$}
\email{haris@akdeniz.edu.tr}
\author{N. Paar$^{1}$}
\email{npaar@phy.hr}
\homepage{http://www.phy.pmf.unizg.hr/~npaar}
\affiliation{$^{1}$Physics Department, Faculty of Science, University of Zagreb, Croatia}
\affiliation{$^{2}$Department of Physics, Akdeniz University, TR-07058, Antalya, Turkey}

\begin{abstract}
{\bf Background:} Inelastic neutrino-nucleus scattering through the weak neutral-current plays
important role in stellar environment where transport of neutrinos determine the
rate of cooling.
Since there are no direct experimental data on neutral-current neutrino-nucleus cross sections
available, only the modeling of these reactions provides the relevant input for
supernova simulations. 
 {\bf Purpose:} To establish fully self-consistent framework for neutral-current neutrino-nucleus
  reactions based on relativistic nuclear energy density functional.
  {\bf Methods:} Neutrino-nucleus cross sections are calculated using weak Hamiltonian 
  and nuclear properties of initial and excited states are
  obtained with relativistic Hartree-Bogoliubov model and relativistic quasiparticle random phase
  approximation that is extended to include pion contributions for unnatural parity transitions.
  {\bf Results:} Inelastic neutral-current neutrino-nucleus cross sections for $^{12}$C, $^{16}$O, 
  $^{56}$Fe, $^{56}$Ni, and even isotopes $^{92-100}$Mo as well as respective cross sections averaged over distribution of supernova neutrinos.
  {\bf Conclusions:}  The present study provides insight into neutrino-nucleus scattering cross sections 
 in the neutral channel, their theoretical uncertainty in view of recently developed microscopic models, and  paves the way for systematic 
 self-consistent large-scale calculations involving open-shell target nuclei.\end{abstract}

\pacs{25.30.Pt, 21.60.Jz, 26.50.+x, 21.30.Fe, 23.40.Bw}
\maketitle

\section{Introduction}

Nuclear weak interaction processes play an important role in the evolution of supernova collapse, e.g., electron capture, beta-decay, and neutrino-nucleus reactions~\cite{Jan.07}. Neutrino-nucleus scattering through the weak neutral current could provide contributions of relevance in stellar environment where transport of neutrinos determine the rate of cooling~\cite{Woo.90,Hax.88,Jan.07}. Recently, inelastic neutrino-nucleus scattering has been introduced in supernova simulations as novel mode of energy exchange between neutrinos and matter
~\cite{Bru.91,Lan.08,Lan.10}. Although this process has no large effect on collapse trajectories, it has significant
contribution to  increasing the neutrino opacities, and it strongly reduces the high-energy tail of the neutrino spectrum emitted in the neutrino burst at shock breakout~\cite{Lan.08}. 
To date, only a single microscopic framework for the neutral-current neutrino-nucleus scattering, based on hybrid model~\cite{Juo.05}, has been included in supernova simulations~\cite{Jan.07}. 
Since the calculations of weak interaction processes in various theoretical
models can result in differences in the reaction rates and cross sections, sometimes larger than
order of magnitude~\cite{Sam.11,Paa.11,Paa.09,Niu.11}, providing insight into the neutral-current 
neutrino-nucleus cross sections from independent models is paramount for assessing 
the impact of the uncertainties in nuclear structure models on the outcomes of 
supernova simulations.

Modeling neutrino-induced reactions is also important in view of studies on modern detectors based on
neutrino scattering on hadrons and nuclei. The ongoing and planned neutrino detector facilities involve variety of  target materials, induced reactions and scientific objectives. These include MOON~\cite{Eji.08}, MiniBooNE~\cite{Agu.09}, NEMO~\cite{Arn.05}, MINOS~\cite{Ada.06}, SNO+~\cite{Chr.09}, OPERA~\cite{Aga.10}, LVD (Large Volume Detector)\cite{Aga.07}, ORLaND experiment at the Spallation Neutron Source (SNS)~\cite{Efr.05}, NOvA neutrino experiment~\cite{Har.05}, etc. In addition
to charged-current neutrino-induced processes in detectors, another possible reaction channel 
includes neutral-current neutrino scattering on hadrons and nuclei, resulting e.g., in small
showers of secondary gamma radiation and creation of electron-positron pairs.  
Although the cross sections in the neutral channel are smaller than in the case of charge-exchange reactions, 
understanding complete detector response necessitates consistent microscopic insight into all relevant processes involved.

Over the past years, several theoretical frameworks have been developed to provide description
of the inelastic neutrino-nucleus scattering in the neutral channel. Due to considerable progress
in the shell model Hamiltonians, a number of neutrino-induced reactions have been described,
also including various decay channels~\cite{Yin.92,Eng.96,Sam.02,Yos.08,Suz.09,Suz.11}. Random phase approximation (RPA)
based on Landau-Migdal force has been employed in calculations of neutrino-induced reaction rates for
r-process nuclei, including those in the neutral channel~\cite{Hek.00,Lan.02}.
The hybrid model combines the shell model for allowed transitions, with the RPA 
to account for the forbidden transitions, allowing systematic calculations for a large number of target nuclei~\cite{Toi.01,Juo.05,Kol.03,Suz.09}.  The Hartree-Fock + RPA based fully on Skyrme functional has been employed in studies of the cross sections for $^{12}$C, $^{16}$O, and $^{208}$Pb~\cite{Jac.99,Jac.02}.
In view of developing neutrino detectors and astrophysical role of neutrino-nucleus reactions,
the relevant nuclear matrix elements have recently been revisited in Ref.~\cite{Cha.09}, and employed
in studies involving target nuclei $^{40}$Ar, $^{56}$Fe, $^{92-100}$Mo and $^{128,130}$Te, based on quasiparticle RPA (QRPA)~\cite{Cha.07,Div.10,Tsa.11,Bal.11}. In another recently developed framework
based on Hartree-Fock-Bogoliubov (HFB) model and QRPA, Brueckner G matrix is employed for two-body interaction by solving the Bethe-Salpeter equation  based on the Bonn CD potential, and pairing correlations have also
been taken into account~\cite{Che.10,Che2.10,Che.11}. Supernova neutrino-$^{56}$Fe cross sections
in the neutral channel have also been explored in the local density approximation taking into account 
the Pauli blocking and Fermi motion effects~\cite{Ath.05}.

Whereas for the charged-current neutrino-nucleus reactions some experimental data are available for $^{12}$C and $^{56}$Fe~\cite{Ath.97,Bod.94,Mas.98,Koe.92}, in the case of neutral-current inelastic neutrino-nucleus scattering there are no experimental data available, except for the ground state transition to the 15.11 MeV state (T=1) in $^{12}$C~\cite{Zei.94,Aue.01}. The indirect experimental insight into the inelastic neutrino-nucleus cross sections can be obtained from inelastic electron scattering. As shown in Ref.~\cite{Lan.04}, magnetic dipole strength distributions for several iron group
nuclei are dominated by isovector Gamow-Teller transitions that can be translated into
inelastic neutral-current neutrino-nucleus cross sections. Due to lack of direct experimental data
on these cross sections, modeling by various approaches is crucial. In this way one can provide 
not only the relevant input for supernova simulations, but also the insight into theoretical 
uncertainties in description of the cross sections in the neutral channel.

In this paper we introduce the framework for the neutral-current neutrino-nucleus inelastic
scattering based on relativistic nuclear energy density functional. Within this framework the nuclear
ground state and various excitations induced in nuclei by the incoming neutrinos are described in a
fully self-consistent approach, i.e., universal effective interaction is employed without 
introducing any adjustments to the specific properties of target nuclei or neutrino energies 
involved.  Although the energy density functional has already been employed in the non-relativistic framework using Skyrme parameterizations~\cite{Jac.99,Jac.02}, the present study provides the first
self-consistent framework to describe neutrino-induced reactions in the neutral channel, involving open-shell target nuclei that necessitate explicit implementation of the pairing correlations.
The relativistic nuclear energy density functional has been successfully employed in studies of giant resonances and exotic modes
of excitation~\cite{Vre.99,Vre.04,PVKC.07,PPP.05,Paa2.09,Kha.11}, $\beta$-decay rates of r-process nuclei~\cite{Nik.05}, muon capture~\cite{Mar.09} and stellar electron capture rates~\cite{Niu.11}, and in constraining the neutron skin in nuclei~\cite{Vre.03,Kli.07}.  In Refs.~\cite{Paa.08,Paa.11} the relativistic proton-neutron QRPA has been employed in
modeling charged-current neutrino-nucleus reactions. In the present analysis of neutrino-induced reactions in the neutral channel, the model necessitates further development of the relativistic QRPA outlined in Ref.~\cite{Paa.03} in order to allow taking into account both natural and unnatural parity excitations in the neutral channel.

The paper is organized as follows. In Sec. II we introduce the basic formalism for the neutrino-nucleus
cross sections in the neutral channel based on weak Hamiltonian and relativistic nuclear energy density functional. The respective
cross sections have been explored in detail for a set of target nuclei in Sec. III. The conclusions of 
the present work are summarized in Sec. IV.

\section{Theoretical background}

The weak process to be considered is inelastic neutral-current neutrino-nucleus reaction,
\begin{equation}
\nu_e + _{Z} X _N  \rightarrow  \nu_e + _{Z}X^*_{N}, 
\end{equation}
where the incoming electron neutrino $(\nu_e)$ scatters on target nucleus $X(Z,N)$ which absorbs part of the neutrino
energy. The interaction between the neutrino
and nucleus is described by weak Hamiltonian, while the properties of initial and final states of
target nucleus are described by effective nuclear interaction, in this particular case formulated
using relativistic energy density functional. The formalism leading
to the expression for the cross section is given in Refs.~\cite{Con.72,Wal.75}. 
The general expression for neutrino-nucleus differential  cross section is derived in terms of relevant 
multipoles of the nuclear weak currents, 
\begin{eqnarray}
& &\left (  \frac{d \sigma_{ \nu  }   }{ d \Omega }   \right )
  =  \frac{ G_F^2 \epsilon^2}{2\pi^2}  
\frac{4\pi}{ 2J_i+1} \nonumber \\
& &\times  \Bigg\{ 
\sum_{J \geq 0 } \Big\{
(1-\hat{\bm{\nu}} \cdot  \bm{\beta} + 2(\hat{\bm{\nu}} \cdot \hat{\bm 
{q}})(\bm{\beta} \cdot \hat{\bm{q}})
\langle J_f || \hat{\mathcal{L}}_J || J_i \rangle | ^2
+ (1+ \hat{\bm{\nu}} \cdot  \bm{\beta}) \langle J_f || \hat{\mathcal 
{M}}_J || J_i \rangle | ^2 \nonumber \\
& & - 2 \hat{\bm{q}} (\hat{\bm{\nu}} +  \bm{\beta}) Re \langle J_f ||  
\hat{\mathcal{L}}_J || J_i \rangle
\langle J_f || \hat{\mathcal{M}}_J|| J_i \rangle^{*} \Big\} \nonumber \\
& &+ \sum_{J \geq 1}
  \Big\{
  {\left(
1-(\hat{\bm{\nu}} \cdot \hat{\bm{q}})(\bm{\beta} \cdot \hat{\bm{q}})
  \right )}
  \left[ \vert \langle J_f || \hat{\mathcal{T}}_J^{MAG} || J_i  
\rangle \vert^2
+ \vert \langle J_f || \hat{\mathcal{T}}_J^{EL} || J_i \rangle  
\vert^2 \ \right ] \nonumber \\
& & + 2\hat{\bm{q}} \cdot (\hat{\bm{\nu}} -  \bm{\beta} )
Re\langle J_f || \hat{\mathcal{T}}_J^{MAG} || J_i \rangle
\langle J_f || \hat{\mathcal{T}}_J^{EL}|| J_i \rangle^{*} \Big\}  
 \Bigg\},
\label{csec}
\end{eqnarray}
where $G_F$ is Fermi constant for the weak interaction and  $\epsilon$ denotes the energy of outgoing neutrino. The momentum transfer $\bm{q}=\bm{\nu}-\bm{k}$ is defined as the
difference between the incoming ($\bm{\nu}$) and outgoing ($\bm{k}$) neutrino momenta,
$\hat{\bm{q}}$ and $\hat{\bm{\nu}}$ denote the corresponding unit vectors, 
and $\bm{\beta} = \bm{k}/ \epsilon$. The transition matrix elements between 
the nuclear initial and final states include transition operators of various multipoles:  
charge $\hat{\mathcal{M}}_J$, longitudinal $\hat{\mathcal{L}}_J$, 
transverse electric $ \hat{\mathcal{T}}_J^{EL}$, and 
transverse magnetic $\hat{\mathcal{T}}_J^{MAG}$ multipole operators, 
expressed in terms of spherical Bessel functions, spherical harmonics, 
and vector spherical harmonics~\cite{Con.72,Wal.75}. Complete calculation
of inelastic neutrino-nucleus scattering necessitates inclusion of a number of multipoles $J$.
Although higher-order multipoles have rather small contributions at low incoming neutrino energies, these can not be neglected at energies about tens of MeV~\cite{Jac.99,Jac.02,Che.11}.
In the present study, multipoles up to $J=5$ contributing to the cross
section in Eq.~(\ref{csec}) will be included in calculations.
In the specific case of the neutrino-nucleus scattering in the neutral channel, the 
transition operators listed above include the following form factors~\cite{Sin.02,Cha.09},
\begin{itemize}
\item vector form factors $F_1^{V}$, $\mu^{V}=F_1^{V}-2MF_2^V$  \\
\begin{equation}
F_1^{V(n)}(q^2) =    -\frac{1}{2} \left ( 1+ \frac{q^2 }{(840 MeV)^2 }   \right )^{-2}
\end{equation}
\begin{equation}
F_1^{V(p)}(q^2) =    \frac{1}{2}(1-4sin^2 \theta_W) \left ( 1+ \frac{q^2 }{(840 MeV)^2 }   \right )^{-2}
\end{equation}
\begin{equation}
\mu^{V(n,p)}(q^2) =   \mu^{V(n,p)}(0)  \left ( 1+ \frac{q^2 }{(840 MeV)^2 }   \right )^{-2}
\end{equation}
%
%
\item axial vector form factor
\begin{equation}
F_A^{(n,p)}(q^2) =    \mp \frac{1}{2} F_A(0) \left ( 1+ \frac{q^2 }{(1032 MeV)^2 }   \right )^{-2}
\label{axialff}
\end{equation}
\item pseudoscalar form factor
\begin{equation}
F_P^{(n,p)}(q^2) =  \frac{2 m_N F_A^{(n,p)}(q^2)}{q^2 +m_{ \pi }^2} \; .
\end{equation}
\end{itemize}
The indices $n,p$ denote the respective form factors for neutrons and protons, $\theta_W$ denotes the 
Weinberg angle, $sin^2\theta_W$=0.2325, and the static values are $F_A(0)$ = -1.2617,  $\mu^{V(n)}(0)$= -1.463, and  $\mu^{V(p)}(0)$=1.054. In the present analysis the
strange quark content in the form factors has been neglected.
By employing the full operator structures in the transition matrix elements, the inelastic neutrino-nucleus cross section is evaluated using Eq.~(\ref{csec}), with an additional quenching factor included in the free-nucleon axial-vector coupling constant, resulting in $F_A(0)$ = 1.0. This quenching corresponds to additional factor 0.8 in $F_A$ (\ref{axialff}). The value of the quenching 
depends on the effective interactions and the model space under consideration, e.g., in the hybrid model quenching factor of 0.74 has been used~\cite{Juo.05}, while 0.8 represents reasonable value for the framework employed in the present study~\cite{Paa.11}.

The transition matrix elements between the initial and final states in Eq.~(\ref{csec}) are determined in a
fully self-consistent framework based on relativistic nuclear energy density functional~\cite{Nik.08,Vre.05}. Therein the nuclear ground state is described with the Relativistic Hartree-Bogoliubov (RHB) model, and excited states are calculated using the relativistic quasiparticle random phase approximation (RQRPA)~\cite{ Paa.03,PVKC.07}.  
The application of relativistic nuclear energy density functional is realized in terms of the self-consistent mean field theory for nucleons and minimal set of meson fields; isoscalar scalar $\sigma$-meson $(J^{\pi}=0^+, T=0)$, isoscalar vector 
$\omega$-meson  $(J^{\pi}=1^-, T=0)$  and the
isovector vector $\rho$-meson $(J^{\pi}=1^-, T=1)$, supplemented with the electromagnetic field. 
The meson-nucleon interaction is included with a minimal set of the interaction terms,
where the vertex functionals include explicit dependence on the vector density.
The details of the RHB model based on this class of effective density-dependent interactions
are given in Ref.~\cite{Nik.02}. For the model parameters that determine the density-dependent 
coupling strength and the meson masses we employ the values of the DD-ME2 parameterization, obtained by simultaneous adjustment of the effective interaction to the binding energies, charge 
radii, differences between radii of neutron and proton density distributions for 12 spherical
nuclei and nuclear matter properties at saturation density~\cite{Lal.05}. The pairing correlations
in open shell nuclei are described by the finite range Gogny interaction, with parameterization
D1S~\cite{Ber.91}.

The RQRPA is formulated in the canonical single-nucleon basis of the RHB 
model, and the residual interaction is derived from the same nuclear energy density functional as
in the RHB model~\cite{Paa.03,NVR.02}. It includes not only configurations composed from 
two-quasiparticle states of positive energy, but also pair-configurations formed from the fully or partially
occupied states of positive energy and empty negative-energy states from the Dirac sea.
In the implementation in modeling the weak interaction processes, the major advantage of
the RHB+RQRPA model is that it is fully consistent in view of the effective interactions
employed. In the particle-hole $(ph)$ and pairing $(pp)$ channels, the same interactions
are used in the RHB equations that determine the canonical quasiparticle basis, and 
in the matrix equations of the RQRPA. In this way, one can employ
the same nuclear energy density functional in description of the weak processes throughout 
the nuclide map without any additional adjustments of the model parameters.

In the present study we further extend the RQRPA framework outlined in Ref.~\cite{Paa.03} by including the pion contributions in order to account both for the natural, $(-1)^J=\pi$, and unnatural, $(-1)^{J+1}=\pi$, parity excitations that take part in inelastic neutrino-nucleus scattering. As shown in Refs.~\cite{Che.10,Che.11}, unnatural parity excitations play an important role in the overall neutrino-nucleus cross sections in the neutral channel, and should be included in a framework aiming to provide consistent and reliable results. Excitations of unnatural parity states necessitate the inclusion of the residual interaction term 
generated by the $\pi$-meson $(J^{\pi}=0^-, T=1)$  exchange. At the Hartree level, i.e. in the RHB, 
the pion does not contribute because it carries unnatural parity and the corresponding mean field breaks parity. The pion major effect comes from the second
and higher order diagrams in the correlated two-pion exchange. The quantum hadrodynamics 
model (QHD II) included in addition to $(\sigma,\omega,\rho)$ meson fields also a pseudoscalar
pion field. However, as pointed out in the RRPA study in Ref.~\cite{Fur.85}, the 
pseudoscalar pion couples too strongly, resulting in total disruption of the ordering of the
lowest excited states. The RRPA analysis showed that implementation of pseudovector pion-nucleon coupling improves the spectrum in comparison to experiment, especially for the pion-like states
$J^{\pi}=0^-,2^-,4^-$.  Therefore, in the present study we use the  pion-nucleon interaction with pseudovector coupling, given in the Lagrangian density as
\begin{equation}
\mathcal{L}^{\textrm{(pv)}}_{\pi} =  \frac{f_{\pi}}{m_{\pi}} \bar{\psi} \gamma_{5} \gamma^{\mu} \partial_{\mu} \vec{\pi} \vec{\tau} \psi
\label{pvcoupling}
\end{equation}
and the propagator for the residual two-body interaction reads,
\begin{equation}
D_{\pi}^{(pv)}(q)=-\frac{1}{q^2+m_{\pi}^2}.
\end{equation}
The standard value for the pseudovector pion-nucleon coupling is ${f_{\pi}^{2}}/{4\pi} = 0.08$,
while the measured pion mass amounts $m_{\pi} = 138$ MeV. Since the one-boson-exchange interaction with pseudovector coupling (\ref{pvcoupling}) contains a contact term, one accounts an additional term for the $\delta$-force to remove its contribution~\cite{PVKC.07}.  The two-body matrix elements of the one-pion exchange interaction and the $\delta$-force in pseudovector coupling are calculated in the momentum space representation according to detailed formalism given in
Ref.~\cite{Ser.01}. When calculating the neutrino-nucleus cross sections in Eq.~(\ref{csec}), 
for each transition operator $\hat{\mathcal{O}}_J$ the matrix elements  
between the ground state and the final state of target nucleus are expressed in terms of
single-particle matrix elements between quasiparticle canonical states, the corresponding occupation factors $v_{\mu}$,$u_{\mu}$ and forward- and backward-going amplitudes $X,Y$, obtained by diagonalization of the RQRPA matrix~\cite{Paa.03},
\begin{eqnarray}
& & \langle J_f || \hat{\mathcal{O}}_J || J_i \rangle  = 
\sum_{\mu\mu'} \bigg\{ X^{ J0}_{\mu\mu'} \langle
\mu || \hat{\mathcal{O}}_J || \mu' \rangle 
 + ~(-1)^{j_{\mu}-j_{\mu'}+J} \, Y^{J0}_{\mu\mu'}
\, \langle \mu' || \hat{\mathcal{O}}_J || \mu \rangle \,
\bigg\} \nonumber \\
 & & \times (u_{\mu}v_{\mu'}+~(-1)^{J}v_{\mu}u_{\mu'}). 
\label{redtrans}
\end{eqnarray}
All relevant transitions between the $|0^+\rangle$ ground state  and $|J_f^{\pm} \rangle$
final states are taken into account in the following calculations.

\section{Results and discussion}

We have employed the framework introduced in Sec. II in modeling the 
neutral-current neutrino-nucleus scattering for a set of target nuclei of
interest for neutrino detector response and understanding the role of
neutrinos in supernova evolution.
In particular, the cross sections have been calculated as a function of the incoming neutrino
energies for $^{12}$C, $^{40}$Ar, $^{56}$Fe,  $^{56}$Ni, and $^{92-100}$Mo isotopes. 
The nuclear matrix elements are obtained using the energy density 
functional with DD-ME2 parameterization~\cite{Lal.05}, supplemented by the Gogny force
D1S to account for the pairing correlations in open shell nuclei~\cite{Ber.91}.
The overall cross section Eq.~(\ref{csec}) includes summation over transitions to
all possible final states characterized by multipoles up to $J=5$ with both positive and
negative parity.

\begin{figure}[!h]
\vspace{-0.6cm}
  \begin{center}
    \centerline{\hbox{\hspace{0.4 cm}
     \includegraphics[angle=270,width=0.8\linewidth]{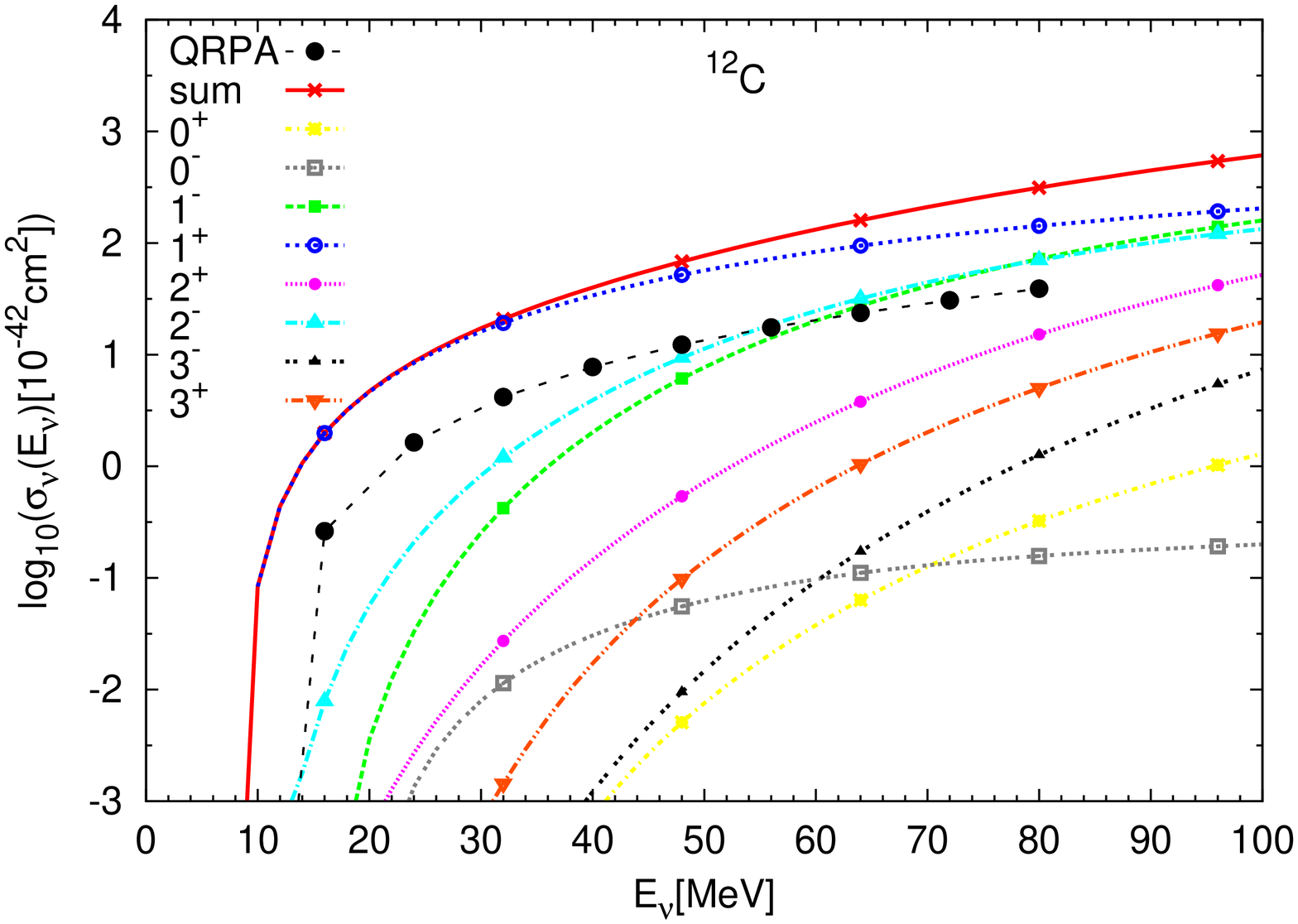}
    }}
    \caption{(Color online) Dependence of the neutrino-nucleus cross sections for the scattering process 
$^{12}$C$(\nu_e,\nu_e')^{12}$C$^*$ on the incoming neutrino energy. The cross sections with separate 
contributions from various multipoles $J_{\pi}=0^{\pm} - 3^{\pm}$ and the total cross sections, 
including $J_\pi=0^\pm - 5^\pm$ states, are shown. The overall cross sections  (stars) are shown in comparison to the QRPA based results (full circles) from Ref.~\cite{Che.10}.}
    \label{fig:ncc12}
  \end{center}
\end{figure}

In Fig.~\ref{fig:ncc12} the calculated electron neutrino-nucleus cross sections are
showed for the inelastic scattering $^{12}$C$(\nu_e,\nu_e')^{12}$C$^*$ for
the range of neutrino energies $E_{\nu}$=0-100 MeV. Complete set of 
multipole states  $J^{\pi}=0^\pm - 5^\pm$ is taken into account in the overall
cross section. Various contributions from the most relevant
multipole states $J^{\pi}=0^\pm - 3^\pm$ to the cross sections are also displayed
separately. For comparison, Fig.~\ref{fig:ncc12} also shows recent results for the overall cross sections  for $^{12}$C target, based on QRPA with Bonn CD potential~\cite{Che.10}.
As one can observe in this figure, at low neutrino energies the overall 
cross sections are dominated by $1^+$ transitions. However, as the energy increases to 100 MeV, the role of other multipoles becomes important, in particular those
of $1^-$, $2^-$ and smaller contribution from $2^+$ states.  In the specific case of neutrino
energy $E_{\nu}$=50 MeV, the present results are at variance with Ref.~\cite{Jac.99},
where multipole contribution from $J^{\pi}=1^-$ dominates over $1^+$. On the
other hand,  the multipole composition
of the cross sections is in qualitative agreement with recent study based on the QRPA~\cite{Che.10}. 
The total cross sections from the present study appear systematically larger than
the respective values obtained using the QRPA~\cite{Che.10}. We turn to this discrepancy
later in discussion of the cross sections.

\begin{figure}[!h]
\vspace{-0.6cm}
  \begin{center}
    \centerline{\hbox{\hspace{0.4 cm}
     \includegraphics[angle=270,width=0.8\linewidth]{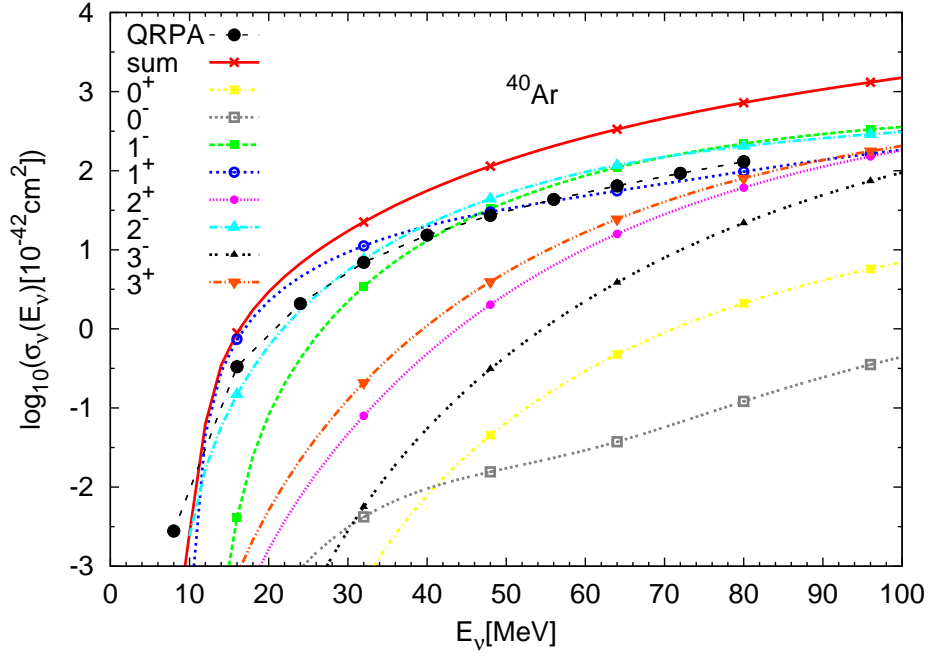}
    }}
    \caption{(Color online)  The same as in Fig.~\ref{fig:ncc12}, but for the scattering process $^{40}$Ar$(\nu_e,\nu_e')^{40}$Ar$^*$. The overall cross sections  (stars) are shown in comparison to the QRPA based results (full circles) from Ref.~\cite{Che.11}.}
    \label{fig:ncar40}
  \end{center}
\end{figure}

In Fig.~\ref{fig:ncar40} the cross sections are shown for the scattering process
$^{40}$Ar$(\nu_e,\nu_e')^{40}$Ar$^*$. In comparison to $^{12}$C, the interplay between
various multipoles becomes more involved. Although at low-energies $1^+$ 
transitions dominate, at $E_{\nu}$ above $\approx$40 MeV, $1^-$ and $2^-$
multipoles have the largest contributions. Neutrino-induced reactions with $^{40}$Ar have
been studied in details in recent work based on QRPA~\cite{Che.11}, in view of
their relevance for detecting core-collapsing supernovae neutrinos. For comparison,
the respective QRPA results from~\cite{Che.11} are also shown in  Fig.~\ref{fig:ncar40}.
The total cross sections from the present analysis appear up to an order of
magnitude larger than the QRPA~\cite{Che.11} ones. Even though
a variety of advanced theoretical frameworks have been developed over the
past years, one can observe considerable theoretical uncertainty inherent
in the modeling of the neutral-current neutrino-nucleus cross sections. These 
uncertainties originate to a large extent to differences in single-particle 
spectra and respective transitions induced by incoming neutrinos. 
In Ref.~\cite{Jac.02}
it has been shown that even within the same model, Hartree-Fock + RPA 
based on Skyrme functional, only small adjustment of single-particle
parameters resulted in $30\%$ increase of the overall neutral-current neutrino-nucleus 
cross sections. Therefore, it is not surprizing that implementation of various models
using independent effective interactions could result in large differences. Consequently, 
it is important to provide reasonable quantitative estimate of theoretical uncertainty in the
cross sections and to critically assess its effect in modeling supernova evolution and
neutrino detector response.
In recent analysis of charged-current neutrino-nucleus cross sections~\cite{Paa.11},
 by employing variety of microscopic models and effective interactions, it has been 
 shown that one can provide reasonable estimate of the theoretical uncertainty 
 in modeling weak interaction processes. 

\begin{figure}[!h]
\vspace{-0.6cm}
  \begin{center}
    \centerline{\hbox{\hspace{0.4 cm}
     \includegraphics[angle=270,width=0.8\linewidth]{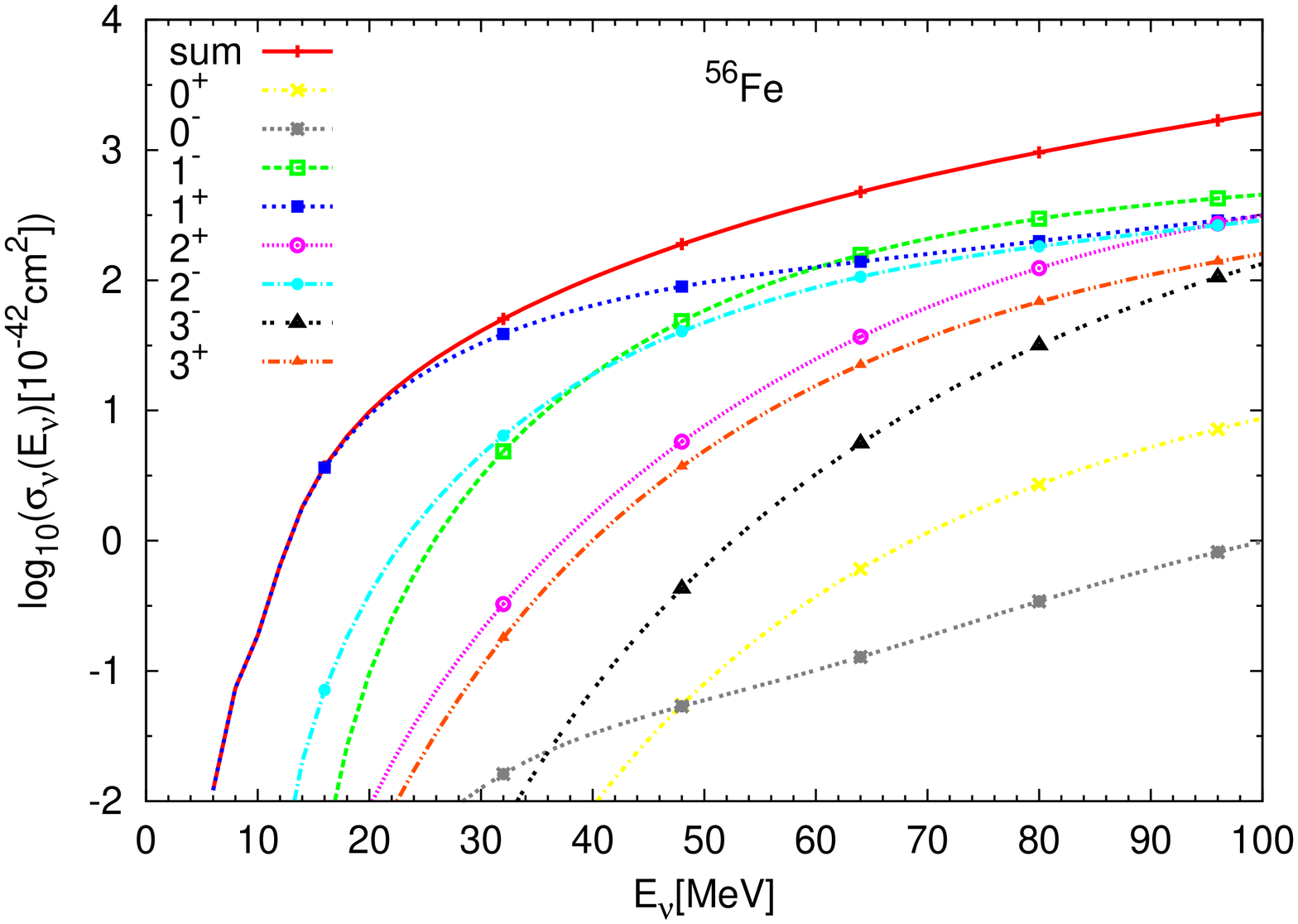}
    }}
    \caption{(Color online) The same as in Fig.~\ref{fig:ncc12}, but for the scattering process  $^{56}$Fe$(\nu_e,\nu_e')^{56}$Fe$^*$.}
    \label{fig:ncfe56}
  \end{center}
\end{figure}

\begin{figure}[!h]
\vspace{-0.6cm}
  \begin{center}
    \centerline{\hbox{\hspace{0.4 cm}
     \includegraphics[angle=270,width=0.8\linewidth]{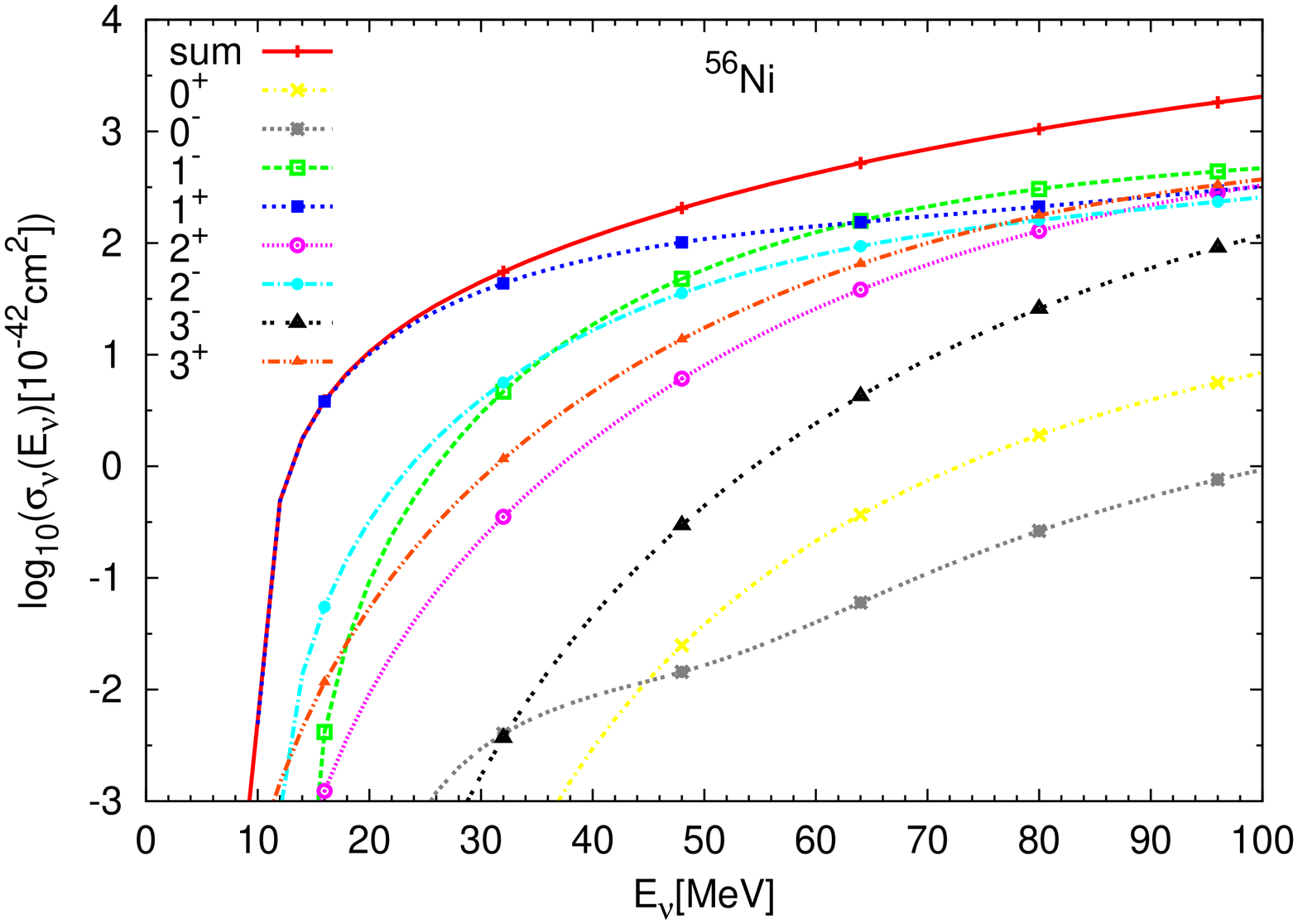}
    }}
    \caption{(Color online) The same as in Fig.~\ref{fig:ncc12}, but for the scattering process $^{56}$Ni$(\nu_e,\nu_e')^{56}$Ni$^*$.}
    \label{fig:ncni56}
  \end{center}
\end{figure}

Figures ~\ref{fig:ncfe56} and ~\ref{fig:ncni56} show the neutral-current 
neutrino-$^{56}$Fe and -$^{56}$Ni cross sections, respectively. Although 
the multipole composition of the cross sections appear in qualitative
agreement, some smaller differences can be noted due to differences in 
neutron and proton numbers and respective excitation spectra. However, 
one can conclude as general property that
$J=1$ states are the most dominant, at lower energies $J^\pi=1^+$ dominates 
while at $E_{\nu} \gtrapprox 65$ MeV transitions $J^\pi=1^-$ have the major contribution.
In addition, at higher energies $J^\pi=2^-$ state also competes with $J^\pi=1^-$ for 
dominance. At high-end neutrino energy $\approx 100$ MeV other multipole transitions
also contribute to the overall cross sections, e.g., $J^\pi=2^\pm, 3^\pm$. 
The calculated cross sections for the scattering process $^{56}$Fe$(\nu_e,\nu_e')^{56}$Fe$^*$ 
are explored in detail in comparison with the hybrid model~\cite{Kol.03} and QRPA based 
framework~\cite{Cha.07}. In Table~\ref{tab:crosscompare}
the RQRPA cross sections are given for a selection of neutrino-energies up to
100 MeV. For comparison, the results are shown both with and without quenching in $F_A$  (\ref{axialff}). It is interesting
to observe that at low neutrino energies, where the cross sections are rather sensitive on
the fine details of the transition spectra, the RQRPA (with quenching) and hybrid  model results 
are in excellent agreement. Although the cross sections from the three models appear in 
the overall qualitative agreement, in the energy region of 
relevance for the supernova neutrino processes ($\approx$20-40 MeV) the RQRPA cross sections 
are up to a factor $\approx$1.5 (2.0) larger than the hybrid model and QRPA results, respectively. It is interesting to note that very recent QRPA study resulted in averaged cross sections for $^{56}$Fe roughly a factor of two larger than for the hybrid model~\cite{Bal.11}.
\begin{table}[!h]
\begin{center}
\begin{tabular}{lcccc}
\hline
\hline
$E_\nu$ [MeV] & w/o quench.& w quench.& Hybrid~\cite{Kol.03}& QRPA~\cite{Cha.07} \\ \hline
10  & 2.91(-1) & 1.87(-1) & 1.91(-1) & 1.01(+0)  \\ 
15  & 4.30(+0) & 2.77(+0) & 2.19(+0) & 2.85(+0)  \\
20  & 1.51(+1) & 9.78(+0) & 6.90(+0) & 5.79(+0)  \\ 
25  & 3.42(+1) & 2.22(+1) & 1.51(+1) & 1.06(+1)  \\ 
30  & 6.26(+1) & 4.08(+1) & 2.85(+1) & 1.87(+1)  \\ 
35  & 1.04(+2) & 6.79(+1) & 4.89(+1) & 3.24(+1)  \\ 
40  & 1.57(+2) & 1.05(+2) & 7.86(+1) & 5.51(+1)  \\ 
45  & 2.32(+2) & 1.54(+2) & 1.19(+2) & 9.05(+1)  \\
50  & 3.24(+2) & 2.16(+2) & 1.72(+2) & 1.43(+2)  \\
55  & 4.38(+2) & 2.94(+2) & 2.39(+2) & 2.15(+2)  \\ 
60  & 5.76(+2) & 3.89(+2) & 3.20(+2) & 3.09(+2)  \\
65  & 7.40(+2) & 5.01(+2) & 4.15(+2) & 4.26(+2)  \\
70  & 9.29(+2) & 6.33(+2) & 5.25(+2) & 5.63(+2)  \\
75  & 1.15(+3) & 7.85(+2) & 6.50(+2) & 7.17(+2)  \\
80  & 1.40(+3) & 9.59(+2) & 7.89(+2) & 8.82(+2)  \\
85  & 1.68(+3) & 1.16(+3) & 9.42(+2) & 1.05(+3)  \\ 
90  & 2.00(+3) & 1.38(+3) & 1.11(+3) & 1.22(+3)  \\ 
95  & 2.36(+3) & 1.63(+3) & 1.29(+3) & 1.38(+3)  \\ 
100  & 2.76(+3) & 1.92(+3) & 1.49(+3) & 1.52(+3)  \\ \hline \hline
\end{tabular}
\end{center}
\caption{The cross sections for ${^{56}}$Fe$(\nu_e,\nu_e'){^{56}}$Fe$^*$ process, given in units of $10^{-42}$ cm$^2$ . The results
of the present analysis without (second column) and with the quenching factor (0.8) in $F_A$ (third column) are compared with the results of the hybrid model~\cite{Kol.03} (forth column) and QRPA based
model from Ref.~\cite{Cha.07} (fifth column).} \label{tab:crosscompare}
\end{table}

\begin{table}[!h]
\begin{center}
\begin{tabular}{lccccc}
\hline
\hline
$E_\nu$ [MeV] & $^{92}$Mo & $^{94}$Mo& $^{96}$Mo& $^{98}$Mo & $^{100}$Mo \\ \hline
10  & 6.32(-2) & 3.58(-1) & 5.25(-1) & 6.67(-1) & 7.82(-1) \\ 
15  & 2.33(+0) & 3.24(+0) & 3.67(+0) & 4.01(+0) & 4.26(+0) \\ 
20  & 9.68(+0) & 1.14(+1) & 1.21(+1) & 1.26(+1) & 1.29(+1) \\ 
25  & 2.47(+1) & 2.76(+1) & 2.86(+1) & 2.95(+1) & 3.00(+1) \\ 
30  & 4.98(+1) & 5.42(+1) & 5.59(+1) & 5.73(+1) & 5.84(+1) \\ 
35  & 8.95(+1) & 9.61(+1) & 9.88(+1) & 1.01(+2) & 1.03(+2) \\ 
40  & 1.46(+2) & 1.55(+2) & 1.59(+2) & 1.63(+2) & 1.67(+2) \\ 
45  & 2.23(+2) & 2.36(+2) & 2.42(+2) & 2.48(+2) & 2.54(+2) \\ 
50  & 3.22(+2) & 3.39(+2) & 3.47(+2) & 3.56(+2) & 3.64(+2) \\
55  & 4.44(+2) & 4.66(+2) & 4.78(+2) & 4.89(+2) & 5.00(+2) \\ 
60  & 5.89(+2) & 6.16(+2) & 6.32(+2) & 6.47(+2) & 6.61(+2) \\ 
65  & 7.58(+2) & 7.91(+2) & 8.10(+2) & 8.30(+2) & 8.47(+2) \\ 
70  & 9.49(+2) & 9.88(+2) & 1.01(+3) & 1.03(+3) & 1.06(+3) \\ 
75  & 1.16(+3) & 1.21(+3) & 1.24(+3) & 1.26(+3) & 1.29(+3) \\ 
80  & 1.40(+3) & 1.45(+3) & 1.48(+3) & 1.51(+3) & 1.54(+3) \\ 
85  & 1.65(+3) & 1.71(+3) & 1.75(+3) & 1.79(+3) & 1.82(+3) \\ 
90  & 1.92(+3) & 1.99(+3) & 2.03(+3) & 2.07(+3) & 2.11(+3) \\ 
95  & 2.22(+3) & 2.29(+3) & 2.34(+3) & 2.38(+3) & 2.42(+3) \\
100  & 2.52(+3) & 2.61(+3) & 2.66(+3) & 2.71(+3) & 2.75(+3) \\ \hline
\end{tabular}
\end{center}
\caption{The total neutral-current neutrino-nucleus cross sections 
for even isotopes $^{92-100}$Mo, given in units of $10^{-42}$ cm$^2$.} \label{tab:mocross}
\label{tableMo}
\end{table}

\begin{figure}[!h]
\vspace{-0.6cm}
  \begin{center}
    \centerline{\hbox{\hspace{0.4 cm}
     \includegraphics[angle=270,width=0.8\linewidth]{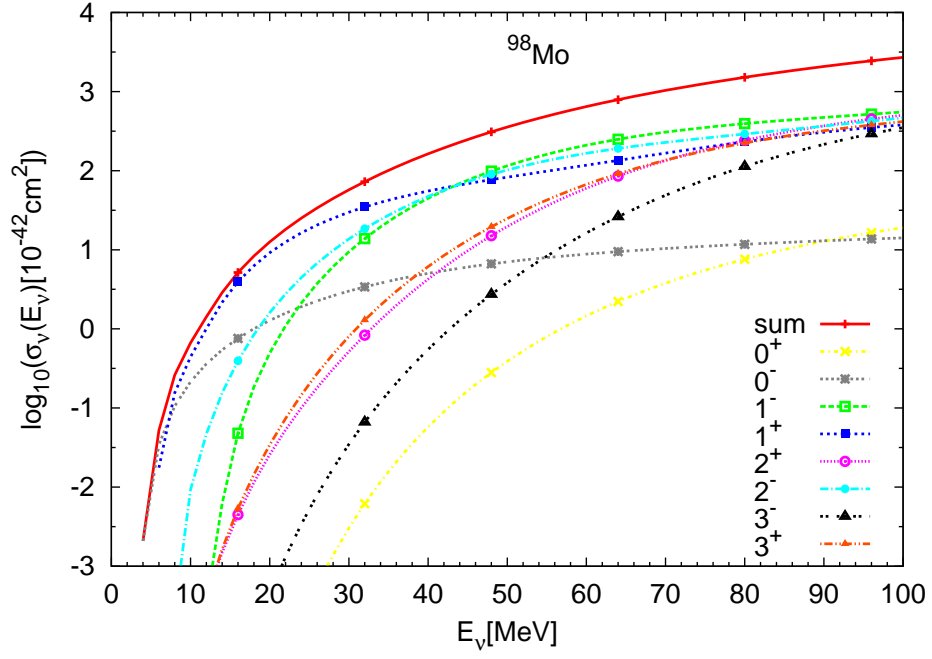}
    }}
     \caption{(Color online) The same as in Fig.~\ref{fig:ncc12}, but for the scattering process
   $^{98}$Mo$(\nu_e,\nu_e')^{98}$Mo$^*$.}
    \label{fig:ncmo98}
  \end{center}
\end{figure}

The scattering cross sections in the neutral channel have also been explored for a set
of Mo isotopes, that recently became interesting due to on-going and future applications of
molybdenum in terrestrial neutrino detectors, MOON~\cite{Eji.08} and NEMO~\cite{Arn.05}, related to neutrino studies and search for the events of neutrinoless double beta decay. 
In the present analysis the cross sections 
have been explored for the most abundant even molybdenum isotopes, $^{92}$Mo, 
$^{94}$Mo, $^{96}$Mo, $^{98}$Mo and $^{100}$Mo. The major contribution in natural
molybdenum comes from $^{98}$Mo, amounting 24.13$\%$. Table~\ref{tableMo} shows the neutral-current neutrino-nucleus cross sections for even isotopes
$^{92-100}$Mo in the range of incoming neutrino energy $E_{\nu_e}$=10-100 MeV. 
By inspecting the numbers, one can observe rather small but systematic increase in 
the cross section values for all neutrino energies.
As can be expected, $^{100}$Mo has the largest cross section of all shown so far simply by 
virtue of large number of active nucleons contributing to the collective nuclear response
in the scattering process. The respective cross sections for $^{98}$Mo and their multipole composition 
are displayed in Fig.~\ref{fig:ncmo98}.
For neutrino energies below 10 MeV 
both $0^-$ and $1^+$ states have relevant contributions. At $E_\nu \approx$ 45 MeV
one can observe the intersection between the main components in the cross
sections: $1^+$ at lower energies and $1^-$, $2^-$ which dominate at higher energies. 
When comparing the overall cross sections to those of other recent studies
based on QRPA~\cite{Bal.11,Che.11}, the present results are systematically larger, but 
within an order of magnitude.

\begin{figure}[!h]
\vspace{-0.6cm}
  \begin{center}
    \centerline{\hbox{\hspace{0.4 cm}
     \includegraphics[angle=0,width=0.8\linewidth]{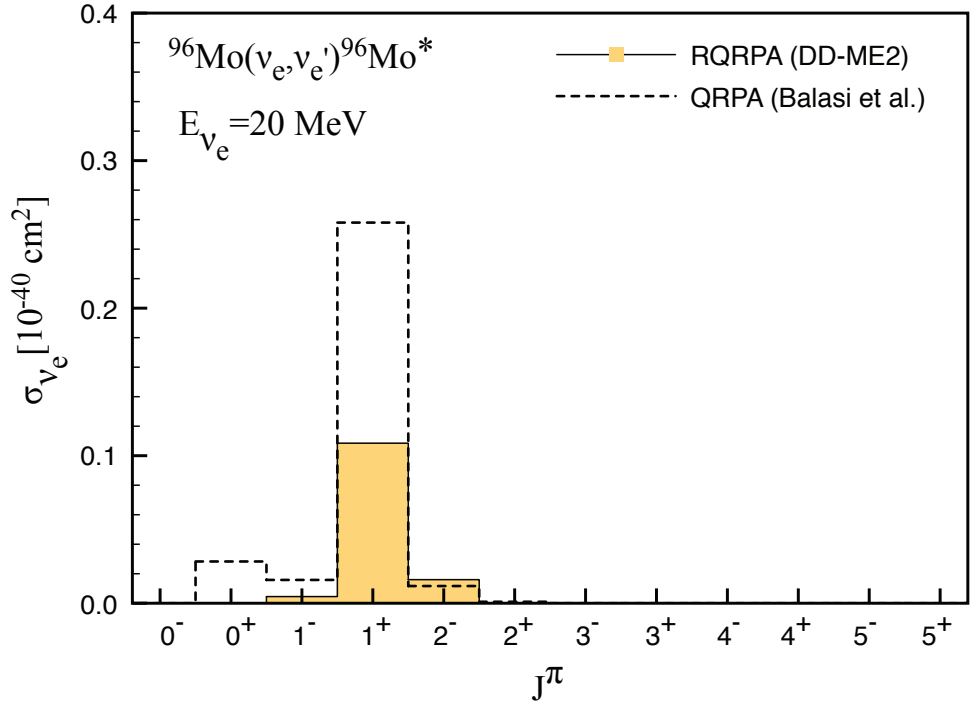}
    }}
     \caption{(Color online)  Contributions of multipole transitions $J^{\pi}=0^{\pm}-5^{\pm}$ in the cross sections for the reaction $^{96}$Mo$(\nu_e,\nu_e')^{96}$Mo$^*$ at incoming electron neutrino energy $E_{\nu_e}$=20 MeV. The results of the present analysis (RQRPA) are compared with QRPA based calculations (Balasi et al., Ref.~\cite{Bal.11}).}
    \label{fig:mo96_20}
  \end{center}
\end{figure}

\begin{figure}[!h]
\vspace{-0.6cm}
  \begin{center}
    \centerline{\hbox{\hspace{0.4 cm}
     \includegraphics[angle=0,width=0.8\linewidth]{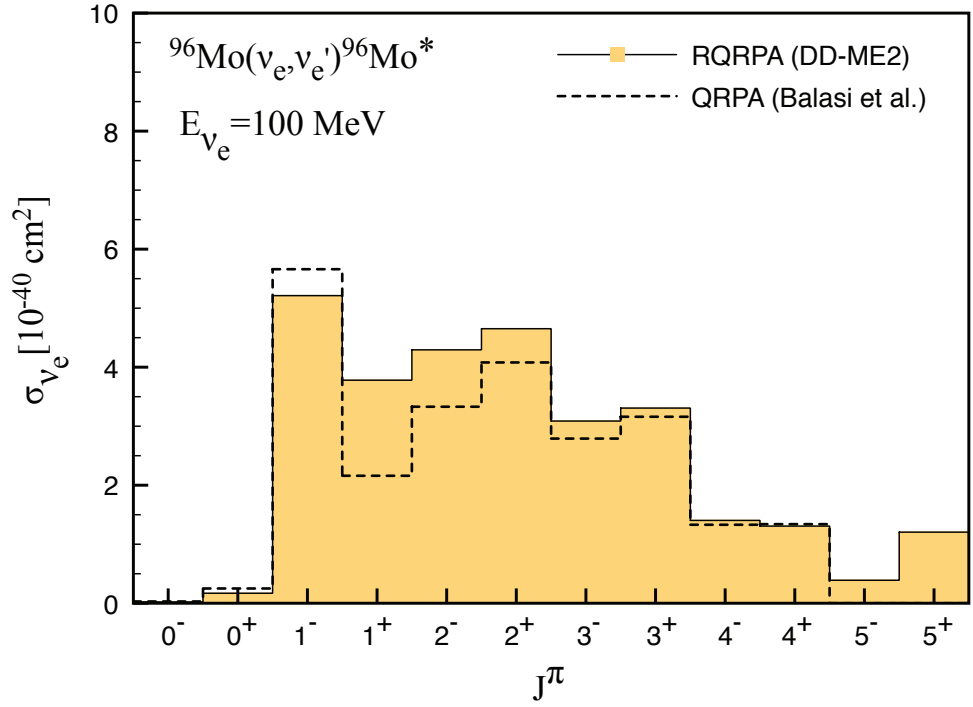}
    }}
     \caption{(Color online) The same as in Fig.~\ref{fig:mo96_20}, but for the neutrino energy $E_{\nu_e}$=100 MeV.}
    \label{fig:mo96_100}
  \end{center}
\end{figure}

In order to explore theoretical uncertainties in modeling neutral-current neutrino-nucleus reactions
in more details, in  Figs.~\ref{fig:mo96_20} and \ref{fig:mo96_100} partial multipole contributions 
to the cross sections for $^{96}$Mo target are shown at incoming electron neutrino
 energies $E_{\nu_e}$=20 and 100 MeV, respectively. The results of the present study (RQRPA) are shown in comparison with the cross sections recently obtained using QRPA (Balasi et al., Ref.~\cite{Bal.11}). Although the total RQRPA cross sections are somewhat larger than those of QRPA, one can observe to a large extent excellent qualitative agreement between the two models based on rather different backgrounds. At low neutrino energy (Fig.~\ref{fig:mo96_20}) in both cases largely dominant excitation channel is $1^+$. The distribution over various multipoles appears rather involved at $E_{\nu_e}$=100 MeV. The main contribution is obtained for $1^-$ transitions, but other multipoles also
 show considerable effects, ranked in the order of importance as follows: $1^-$, $2^+$, $2^-$, $1^+$,  $3^+$, $3^-$, etc. The models based on RQRPA and QRPA result in excellent agreement in relative contributions of various multipoles, except for the anomaly for the QRPA
 $1^+$ channel.

\begin{figure}[!h]
\vspace{-0.6cm}
  \begin{center}
    \centerline{\hbox{\hspace{0.4 cm}
     \includegraphics[angle=270,width=0.8\linewidth]{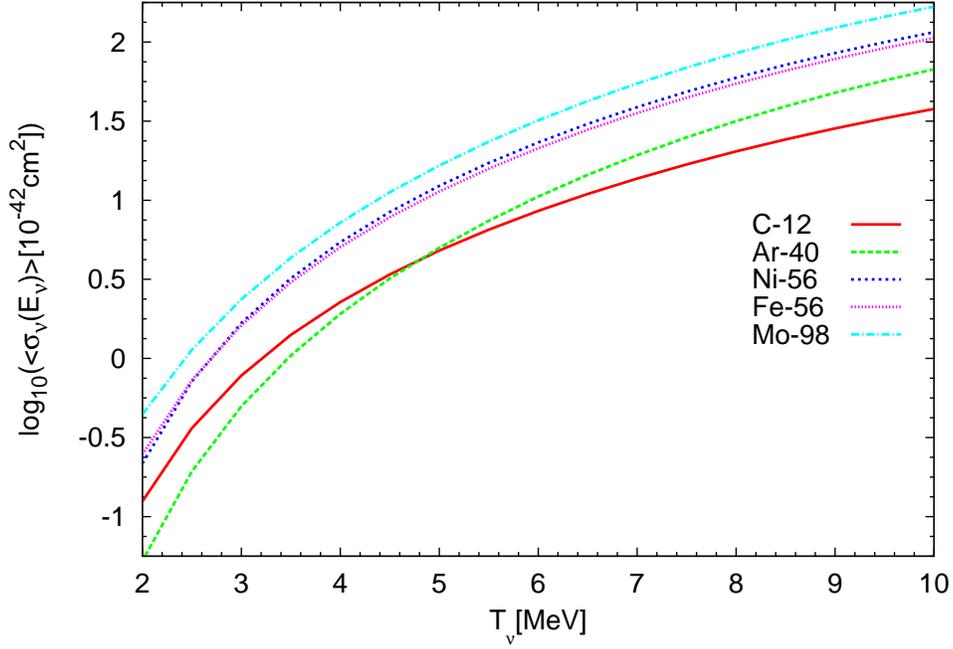}
    }}
    \caption{(Color online) Neutral-current neutrino-nucleus cross sections averaged over the supernova neutrino 
flux in the temperature interval $T_{\nu}=2-10$ MeV. Results are shown for $^{12}$C, $^{40}$Ar, $^{56}$Fe, $^{56}$Ni and $^{98}$Mo target nuclei.}
    \label{fig:alltemp_ax}
  \end{center}
\end{figure}

\begin{figure}[!h]
\vspace{-0.6cm}
  \begin{center}
    \centerline{\hbox{\hspace{0.4 cm}
     \includegraphics[angle=270,width=0.8\linewidth]{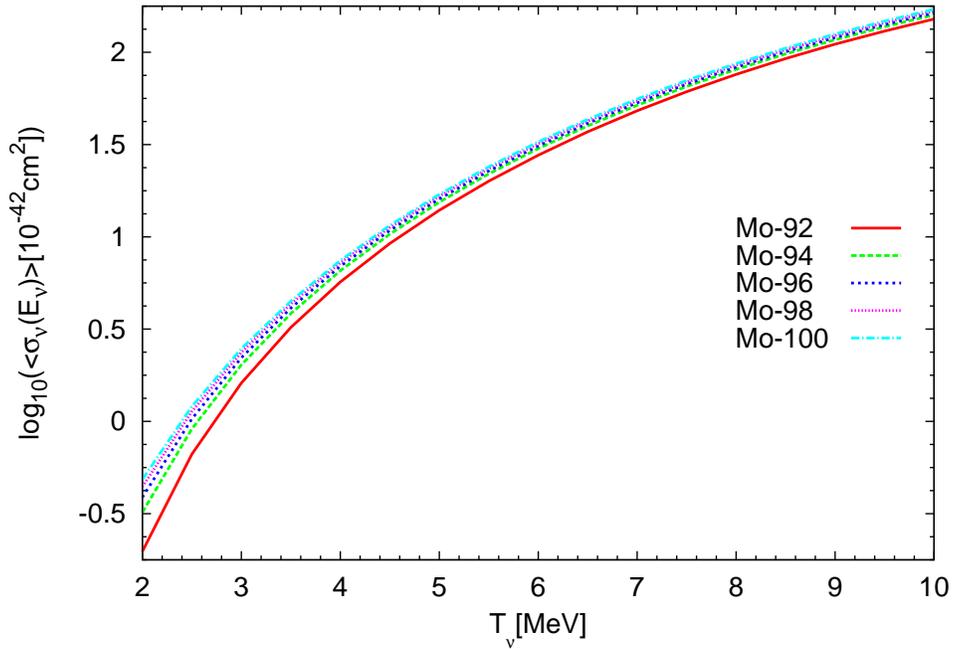}
    }}
    \caption{(Color online) The same as in Fig~\ref{fig:alltemp_ax} but for $^{92,94,96,98,100}$Mo  target nuclei.}
    \label{fig:motemp_ax}
  \end{center}
\end{figure}

An important application of microscopic models of neutrino-nucleus 
reactions is description of the cross sections for stellar neutrinos of relevance
for the neutrino detectors that could provide better insight into fascinating events
in the universe that produce neutrinos.
The calculated cross sections given as functions of the incoming neutrino energy
can be averaged over supernova neutrino flux, that is usually described by the
Fermi-Dirac distribution,
\begin{equation}
f(E_{\nu}) =  \frac{1}{T^3}  \frac{E_{\nu}^2}
{exp \left[ (E_{\nu}/T)-\alpha  \right ] + 1  } \; .
\label{fermidirac}
\end{equation}
Especially interesting is modeling the reaction
rates of neutrinos scattering on nuclei that can be used as targets for the supernova 
neutrino detectors, e.g., $^{40}$Ar, $^{56}$Fe,  $^{56}$Ni, Mo isotopes, etc.
In this way, one can predict expected number of events in detector
that originate from specific stellar environment which determines the production of
low-energy neutrinos. In this work we calculate the neutral-current 
neutrino-nucleus cross sections averaged over the supernova
neutrino flux in the range of temperatures $T_{\nu}= 2 - 10$ MeV, and for
the chemical potential $\alpha$=0. Figure~\ref{fig:alltemp_ax}   shows the respective
flux-averaged cross sections for a set of target nuclei, $^{12}$C, $^{40}$Ar, $^{56}$Fe, $^{56}$Ni and $^{98}$Mo.
As the temperature increases, neutrinos with higher energies have larger contributions
in the averaged cross-sections. The reason is two-fold, i) the Fermi-Dirac distribution shifts
toward higher energies with increased temperature, and ii) the neutrino-nucleus
cross sections increase with neutrino energy and contributions of higher multipole transitions
become significant. In general, for heavier target nuclei the overall cross sections 
are more pronounced. 
In view of the modern neutrino detectors based on molybdenium, 
it is interesting to inspect the results of microscopic calculations for the processes induced
by supernova neutrinos in the most abundant Mo isotopes. In Fig.~\ref{fig:motemp_ax} the respective cross sections, obtained by folding with the Fermi-Dirac distribution (\ref{fermidirac}) for $\alpha$=0, are shown as a function
of temperature for even isotopes  $^{92-100}$Mo. In accordance with the cross sections 
shown in Table~\ref{tableMo}, the averaged cross sections increase with
the number of neutrons in the Mo isotope chain. However, the differences between the averaged 
cross sections are more pronounced at lower temperatures due to larger sensitivity 
of the cross sections to the transitions involved. For example, the ratio $<\sigma(^{100}$Mo$)>~/~<\sigma(^{92}$Mo$)>=$(2.46,1.31,1.18,1.14,1.13) for the set of temperatures T=(2,4,6,8,10) MeV, respectively.

\section{Conclusion}

In summary, modeling of the neutrino-nucleus scattering through the weak neutral current
provides important data for simulations of supernova evolution and detector response to neutrinos emerging from explosive stellar events.
Due to lack of experimental data, it is necessary to provide
independent microscopic insights into the properties of neutrino-induced processes, and assess the theoretical
uncertainty inherent to implementation of various nuclear effective interactions which determine
the transition matrix elements contributing to the neutrino-nucleus cross sections in the neutral
channel.

In this work the self-consistent framework for inelastic
neutral-current neutrino-nucleus scattering is introduced, based on systematic implementation of relativistic nuclear energy density functional with density dependent meson-nucleon couplings. 
The cross sections have been 
formulated using the weak interaction Hamiltonian
and nuclear properties of initial and excited states are obtained by using the RHB+RQRPA, thus allowing studies of open shell target nuclei that necessitate explicit inclusion of the pairing correlations.
In order to include complete set of natural and unnatural parity excited states, the RQRPA residual interaction has been extended using the pion contributions with pseudovector coupling.
In the present analysis, the neutral-current neutrino-nucleus cross sections have been calculated
for the set of target nuclei, $^{12}$C, $^{40}$Ar, $^{56}$Fe, and  $^{56}$Ni. In addition, in view of the MOON~\cite{Eji.08} and NEMO~\cite{Arn.05} experiments based on molybdenum detectors, the present study 
covered the respective neutrino-nucleus cross sections in the neutral channel for the most abundant even isotopes $^{92-100}$Mo. In addition to tables~\ref{tab:crosscompare} and ~\ref{tableMo} presented in this work, complete tables of all calculated cross sections with small step in neutrino energy are available on request.

Comparison of the cross sections and their multipole composition appear to be in reasonable agreement with previous
studies, however, some quantitative differences have been observed. 
From the comparison with calculations based on hybrid model and QRPA, the present analysis provides
an estimate of the theoretical uncertainty in modeling the cross sections in the neutral channel due to implementation
of various theory frameworks and nuclear effective interactions. In the case of ${^{56}}$Fe, it is shown that the overall cross sections exhibit
variations, i.e., at some neutrino energies the cross sections based on RQRPA, QRPA and hybrid model can differ by a factor $\approx$2. This result, together with discrepancies up to an order of magnitude between the RQRPA and QRPA calculations shown for ${^{12}}$C and ${^{40}}$Ar, indicates that in future studies one could critically assess the effect of uncertainties emerging from the calculated cross sections on supernova evolution models or neutrino detector response.

The main advantage of the
present approach to the neutrino-nucleus cross sections in the neutral channel is self-consistent
modeling of all relevant transition matrix elements involving open-shell nuclei, without any additional adjustments of the
model parameters to the nuclear target under consideration. In this way, the present study paves the way for systematic self-consistent large-scale calculations of stellar neutrino-nucleus scattering in the
neutral channel. However, this goal would also necessitate further extension of the model, in order to
include the finite temperature effects in description of nuclei and their excited states in the supernova environment. 
As shown in Refs.~\cite{Sam.02,Juo.05}, at finite temperature the cross sections become somewhat 
enhanced at lower neutrino energies. In the forthcoming study, the present theoretical framework 
will be extended to include finite temperature effects, typical for the supernova environment.

\bigskip
\leftline{\bf ACKNOWLEDGMENTS}
\noindent
This work is supported by MZOS project No.~1191005-1010 and the Croatian Science Foundation.
H.{\DJ}. acknowledges support by the Unity through Knowledge Fund (UKF Grant No. 17/08)
and by TUBITAK project (TUBITAK-TBAG No. 111T275).


\newpage

\end{document}